\documentclass{emulateapj}

\usepackage{natbib}
\usepackage[figuresright]{rotating}
\usepackage{graphicx}
\usepackage{subfigure}
\newcommand{\hei}{He~{\sc i} 2.112 $\mu m$}
\newcommand{\heii}{He~{\sc ii} 2.189 $\mu m$}
\newcommand{\brg}{Br$\gamma$}
\newcommand{\mum}{\ifmmode \mu m \else $\mu m$\fi}

\newcommand{\teff}{\ifmmode T_{\rm eff} \else $T_{\mathrm{eff}}$\fi}
\newcommand{\logg}{\ifmmode \log g \else $\log g$\fi}
\newcommand{\lL}{\ifmmode \log \frac{L}{L_{\odot}} \else $\log \frac{L}{L_{\odot}}$\fi}
\newcommand{\mdot}{$\dot{M}$}
\newcommand{\myr}{M$_{\odot}$ yr$^{-1}$}
\newcommand{\vsini}{$V$ sin$i$}
\newcommand{\vinf}{\ifmmode v_{\infty} else $v_{\infty}$\fi}

\newcommand{\kms}{km s$^{-1}$}
\newcommand{\msun}{\ifmmode M_{\odot} \else M$_{\odot}$\fi}
\newcommand{\zsun}{\ifmmode Z_{\odot} \else Z$_{\odot}$\fi}
\newcommand{\lsun}{\ifmmode L_{\odot} \else L$_{\odot}$\fi}
\newcommand{\rsun}{\ifmmode R_{\odot} \else R$_{\odot}$\fi}

\slugcomment{}

\shorttitle{The nature of S2}
\shortauthors{Martins et al.}

\begin{document}

\title{On the nature of the fast moving star S2 in the Galactic Center \footnote{Based on observations collected at the ESO Very Large Telescope (programs 075.B-0547, 076.B-0259, 077.B-0503, 078.B-0520 and 179.B-0261)}}

\author{F. Martins \altaffilmark{1}, S. Gillessen \altaffilmark{1}, F. Eisenhauer \altaffilmark{1}, R. Genzel \altaffilmark{1,2}, T. Ott \altaffilmark{1}, S. Trippe \altaffilmark{1}}

\email{martins@mpe.mpg.de}


\altaffiltext{1}{Max Planck Institute of Extraterrestrial Physics, Postfach 1312, D-85741, Garching, Germany}
\altaffiltext{2}{Department of Physics, University of California, CA 94720, Berkeley, USA}

\begin{abstract}
We analyze the properties of the star S2 orbiting the supermassive
black hole at the center of the Galaxy. A high quality SINFONI H and K band
spectrum obtained from coadding  23.5 hours of observation
between 2004 and 2007 reveals that S2 is an early B dwarf
(B0--2.5V). Using model atmospheres, we constrain its stellar and wind
properties. We show that S2 is a genuine massive star, and not the
core of a stripped giant star as sometimes speculated to resolve the
problem of star formation so close to the supermassive black hole. We
give an upper limit on its mass loss rate, and show that it is He
enriched, possibly because of the presence of a magnetic field.
\end{abstract}



\section{Introduction}
\label{intro}

The central parsec of our Galaxy hosts a large population of young
massive stars \citep{krabbe95,pgm06}. Their presence is puzzling since
according to standard theories, star formation should not happen so
close to the supermassive black hole SgrA*: the tidal forces are so
large that any molecular cloud should be disrupted before being able
to collapse \citep{morris93}. For most of the young stars, the
solution seems to be star formation in dense accretion disks as
witnessed by the two counter-rotating flat stellar structures around
SgrA* \citep{lb03,genzel03,pgm06}. However, this attractive solution
does not explain the so-called ``S stars'', the group of objects
located within one arcsecond (=0.038 pc) of the black hole. Their
orbits are randomly oriented \citep{frank05} and thus do not fit the
accretion disk scenario. Spectroscopically identified as B stars
\citep{ghez03,frank05}, their true nature as massive stars has been
challenged: they have been proposed to be the hot, luminous cores of
evolved red giant and/or AGB stars the envelope of which has been
stripped by tidal interactions with SgrA* \citep{tal05,kd05}. In that
case, the problem of their formation process vanishes since they might
very well have formed away from the hostile environment of the black
hole before having migrated towards the Galactic Center. Such a
process is too slow to explain the presence of short lived massive
stars close to SgrA*. Hence, these S stars constitute a ``paradox of
youth'' \citep{ghez03}.

In this letter, we analyze the properties of the brightest S-star --
S2 -- and show that it is a genuine massive star.


\section{Observational data and models}
\label{obs_mod}

\begin{figure}
\epsscale{1}
\plotone{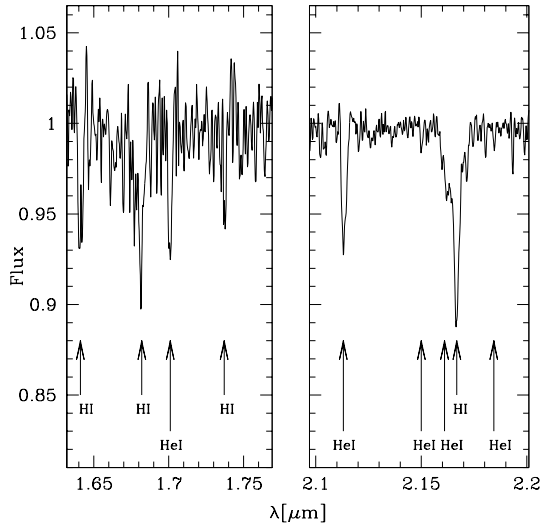}
\caption{Spectrum of S2 after addition of all the SINFONI observations since 2004. The main lines are indicated. \label{spec_s2}}
\end{figure}

To constrain the physical parameters of S2, a high quality spectrum is
needed. Since the observation of the S stars is time consuming, the
S/N ratio obtained in one night is usually limited to 10 at
maximum. To increase this ratio, we have co-added all the S2 spectra
obtained with SINFONI \citep{sinf} in adaptive optics mode (0.0125\arcsec
$\times$0.025\arcsec) on the VLT since 2004. This corresponds to a
total integration time of 23.5 hours in K band. Each spectrum was
carefully extracted by selection of source and background pixels from
which nebular contamination was removed. Note however that this is not
such a critical problem for S2 which has a large radial velocity so
that the position of the stellar absorption core is blueshifted beyond
the nebular emission. We have determined the combined spectrum by a
cross-correlation technique. First, the position of the \brg\ line was
estimated by a simple Gaussian fit to the line. From the such obtained
radial velocities a first combined spectrum was calulated. This
spectrum was then used to cross-correlate all individual spectra
against it. This yielded better estimates for the radial
velocities, which in turn were used to get a more accurate combined
spectrum. This procedure was iterated until the result did not change
anymore. The resulting spectrum shown in Fig.\ \ref{spec_s2} has a S/N
ratio of a few tens in the K band and a resolution of $\sim
4000$. The H band spectrum is noisier due to shorter observation time
and higher extinction.

The main K band lines (\hei\ and \brg) are clearly
identified. But we also detect several weak He~{\sc i} lines at 2.149,
2.161 and 2.184 \mum. Similarly, one He~{\sc i} and three H~{\sc i}
lines are identified for the first time in the H band \footnote{These
lines were seen by \citet{frank05} in the average spectrum of several
stars, but not in individual spectra.}.  We do not see any trace of
\heii. With the current S/N ratio, this line would be detected if it
had a depth of about 1\% of the continuum level.

For the quantitative analysis of this spectrum, we used the atmosphere
code CMFGEN \citep{hm98}. It allows the computation of non-LTE
atmosphere models including winds and line blanketing and is appropriate for hot massive stars. A description of the code is given in \citet{hm98}, and we refer to \citet{gc07} for a summary of
the procedure used to build the models. To derive the parameters of
interest, we have run models with: 19000 $< T_{\rm eff} <$ 30000 K,
3.0 $< \log g <$ 4.75, 0.1 $<$ He/H $<$ 2.0 and 10$^{-8}$ $< \dot{\rm
M} <$ 10$^{-5.5}$ \myr. These models and synthetic spectra
included, in addition to H and He, C, N, O, Si, S and Fe. We adopted the solar
abundances of \citet{gs98}.


\section{The nature of S2: a genuine B star}
\label{param}

\subsection{Spectroscopic classification}

The spectrum shown in Fig.\ \ref{spec_s2} is typical of an early B
star \citep[see also][]{ghez03}. Inspection of the atlases of
\citet{wh97} and \citet{hanson05} shows that for late O stars, \heii\
is expected. We do not detect this line in spite of the good enough
S/N ratio. For stars with spectral types later than B3, the He~{\sc i}
lines vanish. Since we detect several of them, we can safely argue
that S2 has a spectral type between B0 and B2.5. Fig.\ 12 of
\citet{hanson05} shows that in supergiants and giants, \brg\ is
clearly separated from He~{\sc i} 2.161 \mum, while for dwarfs \brg\
is broad enough to merge with the He~{\sc i} line. This latter
morphology is similar to what we observe for S2 (see below for a
quantitative comparison). Hence, in addition to being
spectroscopically identified as an early B star, we can conclude from
these qualitative arguments that \textit{S2 is also a dwarf and not a
supergiant/giant}.

\subsection{A mass estimate}

The key question we want to answer here is the exact nature
of S2. Spectroscopically, S2 is unambiguously an early B dwarf.
However, this does not necessarilly mean that S2 is a young,
massive star. It could be the core of an older, evolved star which
would have lost its envelope through tidal interaction with SgrA* but
would still spectroscopically look like a B star. In that case, the
star could have formed far away from SgrA* before being dragged to its
proximity and experiencing an envelope stripping. There would be no
paradox of youth.

Recently, \citet{kd05} and \citet{dray06} modeled this stripping
process as well as the subsequent evolution of the remaining core in
the vicinity of SgrA*. They showed that, under certain condition on
the IMF and the capture rate, the population of S stars could be
explained by tidal stripping. The argument is however based on the
number of stars predicted at the position of the S stars in the HR
diagram. Expressed differently, the conclusion comes from the fact
that the modeled stripped stars can reach the luminosities and
effective temperatures of B stars. But strictly speaking, this does
not exclude that the S stars are genuine massive stars, which would
lie at the same position in the HR diagram.

For that, the only parameter which can unambiguously be used is the
stellar mass. The most massive stars able to experience envelope
stripping are AGB stars. According to stellar evolution, such objects
are the evolved descendents of main sequence stars with M $<$ 8
\msun. The core of such objects is actually much less massive: Table 4
of \citet{fc97} shows that their mass is lower than 1 \msun. In the
scenario explored by \citet{dray06}, only stars having lost at least
99\% of their envelope could explain the S stars. Thus, their mass
should be at most $\sim$ 1 \msun.

To estimate the mass of S2, we can rely on gravity and radius: $M = g
R^{2}/G$ where $g$ is the gravity, $R$ the radius and $G$ the
gravitational constant. Gravity can be derived from the shape of \brg,
while radius is straightforwardly obtained from the knowledge of
effective temperature and luminosity. Below, we explain how we
proceeded to constrain these parameters.

\begin{table}
\begin{center}
\caption{Physical properties derived from our quantitative analysis for different \teff\ appropriate for B0--2.5V stars.\label{tab_res}}
\begin{tabular}{l|rrrrr}
\tableline
\teff\ [K]             & 19000 & 22000 & 25000 & 27000 & 30000 \\
\hline                 & & & & \\
\lL\                   & 4.30 & 4.45 & 4.60 & 4.65 & 4.80 \\
\logg\                 & 3.80 & 4.00 & 4.00 & 4.00 & 4.25 \\
$\log g_{min}$         & 3.55 & 3.72 & 3.77 & 3.89 & 3.86 \\
R [\rsun]              & 13.1 & 11.6 & 10.7 & 9.7  & 9.4  \\
R$_{min}$ [\rsun]      & 11.4 & 10.1 & 9.3  & 8.5  & 7.3  \\
M$_{min}$ [\msun]      & 16.8 & 19.5 & 18.6 & 20.5 & 14.1 \\
He/H [\#]              & 1.20 & 0.50 & 0.45 & 0.55 & 0.80 \\
He/H$_{min}$ [\#]      & 0.85 & 0.25 & 0.25 & 0.30 & 0.30 \\
\mdot\ [$10^{-7}$\myr] & $<$3 & $<$3 & $<$3 & $<3$ & $<$3 \\
\tableline
\end{tabular}
\end{center}
\end{table}

\textit{Effective temperature}. \teff\ is usually derived from the
ratio of lines from two consecutive ionization states: He~{\sc i} /
He~{\sc ii} lines for O stars and Si~{\sc iii} / Si~{\sc iv} for early
B stars. Unfortunately, S2 does not show He~{\sc ii} and Si lines in
the K band. Hence, we can only rely on semi-quantitative arguments to
estimate \teff. We have shown above that S2 is a B0V to B2.5V
star. According to the recent studies of \citet{dufton06} and
\citet{trundle07}, such stars have 19000 $<$ \teff\ $<$ 30000 K. We
have thus run models for \teff\ = 19, 22, 25, 27 and 30 kK.

\textit{Luminosity}. To estimate the star's luminosity, for each \teff\, we adjusted \lL\ in our models to match the absolute K band magnitude of S2. This magnitude is defined as 
$MK=mK-A_{K}-DM$
with $mK=14.0$ the observed magnitude of S2 \citep{pgm06},
$A_{K}=2.25$ the K band extinction at the position of S2
\citep{schoedel07} and $DM=14.50$ the distance modulus for a distance
to the Galactic Center of 7.94 kpc \citep{frank03a}. The resulting
absolute K magnitude of S2 is -2.75. The range of values we obtain for
\lL\ is 4.30--4.80. In practice, the uncertainty of $\pm$0.2 on
$A_{K}$ and 0.5 kpc on the distance to the Galactic Center translate
into an uncertainty of about 0.12 in \lL.

\textit{Radius}. The radius of the star is simply derived from $L = 4\pi \sigma_{B} R^{2} T_{eff}^{4}$ ($\sigma_{B}$ being the Boltzmann constant). The minimum radius allowed by our study (corresponding also to the minimum mass of the star -- see below) is estimated for a luminosity reduced by the uncertainty (0.12 dex). The values of $R_{min}$ are given in Table \ref{tab_res}.

\begin{figure}
\epsscale{1}
\plotone{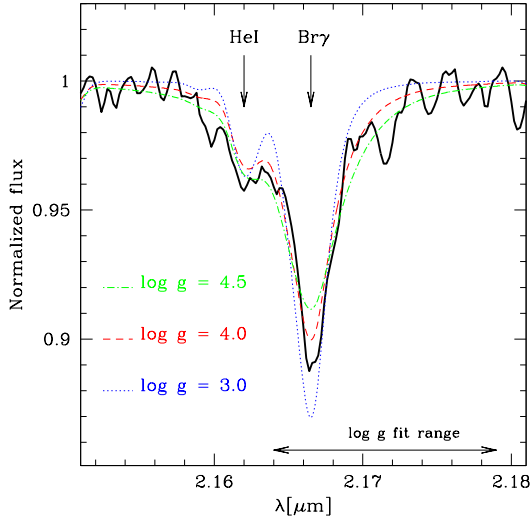}
\caption{Determination of \logg\ for the case \teff\ = 22000 K. The black solid line is the observed \brg\ line. The dashed (dotted, dot-dashed) line is a model with \logg\ = 4.0 (3.00, 4.50). All other parameters are kept constant (especially He/H=0.5). The wavelength range used for the \logg\ determination is indicated. \label{s2_logg_t22}}
\end{figure}

\begin{figure}
\epsscale{1}
\plotone{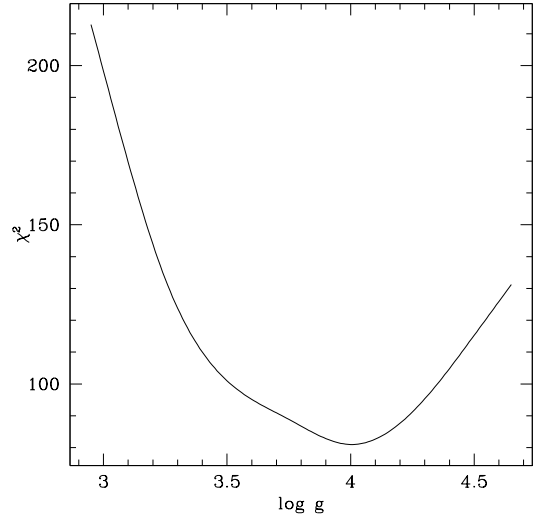}
\caption{$\chi^{2}$ as a function of \logg\ for the \teff\ = 22000 K model. The best fit model has \logg\ = 4.0, as also seen from Fig.\ \ref{s2_logg_t22}. \label{chi2_logg_t22}}
\end{figure}

\textit{Gravity}. \logg\ was constrained from the shape
of \brg. \citet{repolust05} showed that its wings (resp. absorption
core) get broader (resp. weaker) when gravity increases. For large
\logg, the blue wing merges with the He~{\sc i} 2.161\mum\ line as
discussed previously. Fig.\ \ref{s2_logg_t22} shows how the line
profile changes when \logg\ increases from 3.0 to 4.5 for \teff\ =
22000 K. The best models correspond to \logg\ $\sim$ 4.0. A
quantification of the goodness of the fit in the \brg\ region of the
spectrum was performed by means of a $\chi^{2}$ analysis. To minimize
contamination by HeI, we used only the range 2.164-2.179\mum. Fig.\
\ref{chi2_logg_t22} shows the result of this analysis and confirms
that \logg\ = 4.0 is preferred. From this curve, we also derive a 3
$\sigma$ lower limit to \logg\ ($\log g_{min}$).  Table \ref{tab_res}
shows the results for other \teff. The values of \logg\ we obtain are
all typical of a dwarf star \citep[\logg $\sim$ 4.0,][]{trundle07},
confirming our previous ``spectroscopic'' findings.

\textit{Rotational velocity}. We derived a projected rotational
velocity of 100$\pm$30 \kms. It best accounts for the shape of the
\hei\ doublet: for larger \vsini\ the lines merge into a single
component, for lower values, they are too seperated. This value is
commonly found for early B stars: according to \citet{abt02}, \vsini\
= 127$\pm$8 (108$\pm$8) \kms\ for B0--2V (B3--5V) stars.

Table \ref{tab_res} summarizes, for each \teff, the values of \logg\
and radius. Since we want to test if S2 is the core of an AGB star, we
also give the minimum mass of S2. All our estimates are larger than
14.1 \msun. This is larger than the initial mass of AGB stars, and
consequently is well above the mass of the core of such a star. From
that, one can thus safely conclude that \textit{S2 is not a stripped
AGB star, and is really a genuine B star}.


\section{Mass loss rate and Helium content}
\label{param}

\subsection{Mass loss rate}

\begin{figure}
\epsscale{1}
\plotone{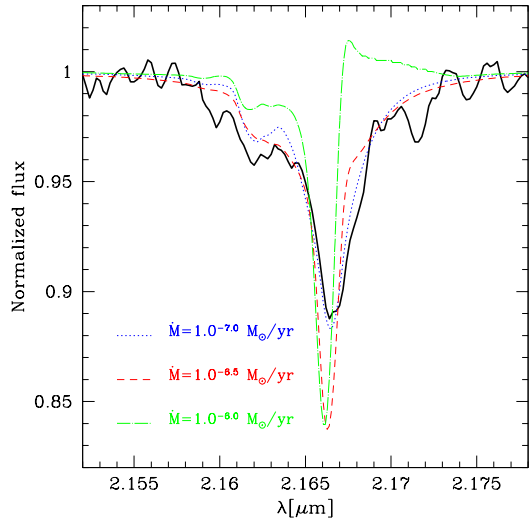}
\caption{Effect of mass loss on \brg. When \mdot\ increases, \brg\ starts to depart from a pure absorption to an emission profile. Characterizing the value of \mdot\ for which this occurs provides an upper limit on the mass loss rate of S2. \label{s2_mdot_t22}}
\end{figure}

Below a certain threshold, \brg\ is insensitive to \mdot. When \mdot\
increases above this limit, the \brg\ absorption profile is
progressively filled by emission. By determining this threshold, we
can set an upper limit on the mass loss rate of S2 (see Fig.\
\ref{s2_mdot_t22}). In practice, the quantity we constrain is the wind
\textit{density} $\rho = \dot{M}/(4 \pi R^{2} v_{\infty})$ (where $R$
is the stellar radius and $v_{\infty}$ the terminal
velocity). Deriving a mass loss rates implies we know
$v_{\infty}$. Since the lines we observe in the S2 K band spectrum are
insensitive to $v_{\infty}$, we have to assume values. Terminal
velocities of B dwarfs are difficult to estimate since the winds are
usually weak and the spectra do not show P-Cygni lines or emission
lines from which it is usually derived. Hence, we have adopted
$v_{\infty}$ = 1000 \kms\ which is appropriate for late O dwarfs
\citep[e.g.][]{jc03}. If $v_{\infty}$ would be lower than 1000
\kms, estimated mass loss rates would be lower. Our assumptions thus
lead to conservative upper limits of \mdot\ $\lesssim 3\times 10^{-7}$
\myr. This is lower than \mdot\ required by \citet{loeb04} to feed
SgrA* without transport of angular momentum.

\subsection{Helium content}

\begin{figure}
\epsscale{1}
\plotone{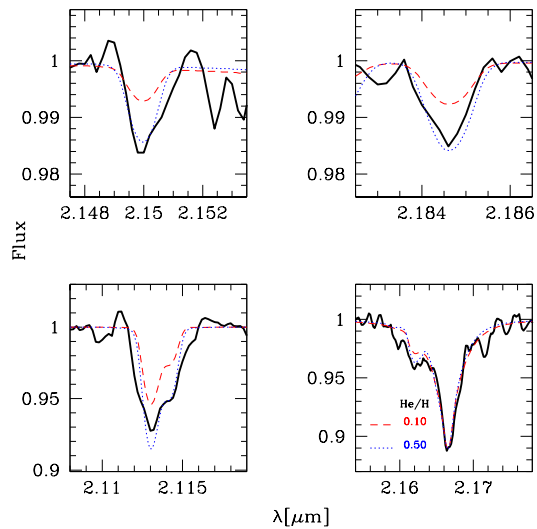}
\caption{Determination of He/H. The black solid line is the observed spectrum of S2. The red dashed (blue dotted) line is a model with He/H=0.10 (0.50). \teff\ = 22000 K and \logg = 4.0. The He rich model is required to fit the S2 spectrum. \label{s2_he}}
\end{figure}

The near infrared spectrum of S2 is dominated by H and He lines. We
thus could constrain the He/H abundance ratio. Qualitatively, when
this ratio increases, all He~{\sc i} lines have stronger absorption
profiles, while H~{\sc i} lines get weaker. In practice, we used the
He~{\sc i} 2.149 $\mu m$ and He~{\sc i} 2.184 $\mu m$ lines to
constrain He/H since these lines are almost insensitive to
microturbulence, which is not the case of \hei\ and He~{\sc i} 2.161
$\mu m$. Fig.\ \ref{s2_he} shows the changes in the He~{\sc i} line
profiles when He/H is varied between 0.1 and 0.5 (model with \teff\ =
22000 K). We find that the He/H ratio varies from about 0.45 at \teff\
= 25000 K up to as high as 1.20 at \teff\ = 19000 K. We proceeded as
for \logg\ ($\chi^{2}$ analysis) to derive a 3 $\sigma$ lower limit on
the value of He/H. All these lower limits are larger than 0.25,
confirming that S2 is He enriched.

With He/H $>$ 0.25, S2 falls into the category of the so-called ``He
rich'' stars, a class of B stars with He/H between 0.3
and 10 \citep{smith96}. Interestingly, these peculiar stars have a
very narrow distribution of spectral types centered around B2, similar
to what we find for S2 (B0--2.5V). The best studied He-rich star is
$\sigma$ OriE. The origin of its surface abundance pattern is
explained by a combination of specific wind properties and magnetic
field. The star should have a wind weak enough for ion decoupling to
occur, i.e. the radiative acceleration, essentially due to metals, is
not redistributed among the passive plasma (mainly H and He) because
of a too low density and reduced Coulomb forces \citep{kk01}. In that
case, Helium might accumulate at the surface of the star. But this
extra Helium can remain on the surface only if turbulence, and its
associated mixing effect, is suppressed. This calls for a magnetic
field strong enough to freeze the stellar surface
\citep{hg99}. Our upper limit on the mass loss rate of S2 is
consistent with a relatively weak wind. In this scenario, chemical
inhomogeneities (spots) are expected on the surface as a result of the
interplay between the magnetic field geometry and the wind. We cannot
test this prediction with our observations since we do not spatially
resolve the stellar surface of S2. Similarly, no CNO abnormalities are
expected in this model, which we cannot test either due to the absence
of CNO lines (observed and expected) in the near-IR spectrum of S2.

If the wind is strong enough to develop into a normal homogenous
outflow, one might still observe a significant He enrichment if the
star has a magnetic field and rotates fast enough. Indeed
\citet{mm05} predict that a strong (10$^{4}$ G) magnetic field
can lead to solid body rotation of the star. This favors the diffusion
of species by meridional circulation. As a consequence, a star can
experience strong surface He enrichment: Fig.\ 10 of \citet{mm05}
shows that a 15 \msun\ star can reach a He mass
fraction of 0.3 (around 0.12 in He/H) in 12 Myr. Contrary to the
previous scenario, large N and reduced C abundances are also expected.

In view of the available models and of the currently known stellar
properties, we thus tentatively propose that S2 is a magnetic star.

\acknowledgments

We thank John Hillier for his help with CMFGEN, as well as Corinne
Charbonnel and Ana Palacios for interesting discussions on AGB
stars. We thank Andrea Ghez and an anonymous referee for useful
comments.



\end{document}